\documentclass[conference,10pt]{IEEEtran}
\makeatletter
\def\ps@headings{%
\def\@oddhead{\mbox{}\scriptsize\rightmark \hfil \thepage}%
\def\@evenhead{\scriptsize\thepage \hfil \leftmark\mbox{}}%
\def\@oddfoot{}%
\def\@evenfoot{}}
\makeatother
\usepackage{graphicx}
\usepackage{balance}
\usepackage{subfigure}
\usepackage{amsmath,amsfonts}
\usepackage[belowskip=-8pt,aboveskip=-1pt]{caption}

\usepackage{color}

\begin{document}
\title{Distributed MAC and Rate Adaptation for Ultrasonically Networked Implantable Sensors}

\author{\IEEEauthorblockN{G. Enrico Santagati and Tommaso Melodia}
\IEEEauthorblockA{State University of New York at Buffalo, NY, USA\\
\{santagat, tmelodia\}@buffalo.edu}
\and
\IEEEauthorblockN{Laura Galluccio and Sergio Palazzo}
\IEEEauthorblockA{University of Catania, Italy\\
\{laura.galluccio, sergio.palazzo\}@dieei.unict.it}}

\maketitle

\begin{abstract}
The use of miniaturized biomedical devices implanted in the human body and wirelessly internetworked 
 is promising a significant leap forward in medical treatment of many pervasive diseases.  Recognizing the well-understood limitations of traditional radio-frequency wireless communications in interconnecting devices within the human body, in this paper we propose for the first time to develop network protocols for implantable devices based on ultrasonic transmissions. We start off by assessing the feasibility of using ultrasonic propagation in human body tissues and by deriving  an accurate channel model for ultrasonic intra-body communications. Then, we propose a new ultrasonic transmission and multiple access technique, which we refer to as Ultrasonic WideBand (UsWB). UsWB is based on the idea of transmitting information bits spread over very short pulses following a time-hopping pattern. The short impulse duration results in limited reflection and scattering effects, and its low duty cycle reduces the thermal and mechanical effects, which are detrimental for human health.  We then develop a multiple access technique with distributed control to enable efficient simultaneous access by interfering devices  based on minimal and localized information exchange and on measurements at the receiver only. 
Finally, we demonstrate the performance of UsWB  through a multi-scale simulator that models the proposed communication system at the acoustic wave level, at the physical (bit) level, and at the network (packet) level. 

\end{abstract}

\IEEEpeerreviewmaketitle

\section{Introduction} \label{sec:intro}
Body area networks have received considerable attention in the last five years \cite{survey, CodeBlue}, driven by the fascinating promise of a future where carefully engineered miniaturized biomedical devices implanted, ingested or worn by humans can be wirelessly internetworked to collect diagnostic information  (e.g., measuring the level of glucose in the blood of diabetic patients) and to fine-tune medical treatments (e.g., adaptively regulate the dose of insulin administered) over extended periods of time \cite{wons, Hogg}.
  
Yet, most research to date has focused on communications {\it along the body surface} among devices using traditional electromagnetic radio-frequency (RF) carrier waves, while the underlying challenge of enabling networked {\it intra-body} miniaturized (at the micro or nano scale) sensors and actuators that communicate through body tissues is substantially unaddressed. The main obstacle to enabling this vision of networked implantable devices is posed by the physical nature of propagation in the human body, which is composed primarily (65\%) of water, a medium through which RF electromagnetic waves notoriously hardly propagate, even at relatively low frequencies. Accordingly, most research has focused on 
reducing the radiated power to avoid overheating of tissues \cite{ieee}.

These formidable challenges cannot be overcome unless a major paradigm shift in networking through body tissues is made to address the limitations of RF propagation in the human body. Therefore, in this paper {\it we present the first study proposing the use of ultrasonic waves to wirelessly internetwork in-body devices.}

Acoustic waves, typically generated through piezoelectric materials, are known to propagate better than their RF counterpart in media composed mainly of water. Since World War II, piezoelectrically generated acoustic waves have found application, among others, in underwater communications (typically at frequencies between 0 and 100 kHz \cite{BasagniBP04}), in indoor localization in sensor networks \cite{spiderbat}, and, massively, in ultrasonic medical imaging \cite{ultras_med2}. While communication at low frequencies requires sizable transducers, innovations in piezoelectric materials and fabrication methods, primarily driven by the need for resolution in medical imaging, have made miniaturized transducers, at the micro \cite{transd, transd2} and even nano scales \cite{nanotransd} a reality. Moreover, the medical experience of the last decades has demonstrated that ultrasound is fundamentally safe, as long as acoustic power dissipation in tissue is limited to specific safety levels \cite{Hogg, wons}. It is also known that ultrasonic wave heat dissipation in tissues is minimal compared to RF waves \cite{Cheung84}.

We lay our foundation on the fundamental physics of ultrasound propagation and move up layers of the protocol stack with the following core contributions:\\
{\bf $-$ Feasibility of Ultrasonic Communications in the Human Body.} We start off by discussing the fundamentals of ultrasonic propagation in tissues, and by exploring important tradeoffs, including the choice of a transmission frequency, transmission power, bandwidth, and transducer size. \\
{\bf $-$ Ultrasonic Channel Modeling. }  We derive a model of the channel impulse respons for ultrasonic communications in tissues and observe that the inhomogeneity of the human body, characterized by a multitude of very small organs, causes severe multipath effect, thus making detection and decoding a challenging operation. In addition, we observe that the low speed of sound in tissues leads to high delays that need to be considered in system design.\\
{\bf $-$ Ultrasonic Wideband Design.} To address the severe effect of multipath, we design and propose a new ultrasonic transmission and multiple access technique, which we refer to as Ultrasonic WideBand (UsWB). Ultrasonic wideband is based on the idea of transmitting very short pulses following an adaptive time-hopping pattern. The short impulse duration results in limited reflection and scattering effects, and its low duty cycle reduces thermal and mechanical effects that are detrimental for human health. A variable length spreading code is superimposed to the time-hopping pattern to further combat the effect of multipath and to introduce waveform diversity among interfering nodes. \\
{\bf $-$ Ultrasonic Wideband Adaptation and Multiple Access Control}. We then study and develop multiple access techniques with distributed control to enable multiple access among interfering implantable devices.  
 Our proposed scheme is based on the idea of regulating the data rate of each transmitter to adapt to the current level of interference by distributively optimizing the code length and duration of the time hopping frame. \\ 
 {\bf $-$ System Performance Evaluation.} We evaluate the proposed scheme through a multi-scale simulator that evaluates our system at three different levels, i.e., (i) at the wave level by modeling ultrasonic propagation through reflectors and scatterers, (ii) 
 at the bit level by simulating in detail the proposed ultrasonic transmission scheme, (iii) at the packet level by simulating networked operations and distributed control and adaptation to evaluate metrics such as network throughput and packet drop rate.

The rest of the paper is organized as follows.
%In Section \ref{related} we discuss relevant literature in the field.
In Sections \ref{propagation} and \ref{sec:channel} we discuss fundamental aspects of ultrasonic physical propagation
in human tissues and channel modeling.
In Section \ref{mac} we outline the design of UsWB and in Section \ref{adapt} we illustrate the proposed medium access control protocol.
Performance evaluation results are discussed in  Section \ref{performanceevaluation}.
Finally, in Section \ref{conclusions} we conclude the paper.% and discuss future work.

 %%%%%%%%%%%%%%%%%%%%%%%%%%%%

\section{Ultrasonic Propagation in Human Tissues } \label{propagation}

Ultrasonic waves originate from the propagation of mechanical vibrations of particles in an elastic medium at frequencies above the upper limit for human hearing, i.e., $20\:\mathrm{kHz}$. Even if each particle oscillates around its rest position, the vibration energy propagates as a wave traveling from particle to particle through the material. 
Acoustic propagation through a medium is governed by the acoustic wave equation (referred to as the Helmhotz equation), which describes pressure variation over the three dimensions, $\nabla^2 {P} - \frac{1} {c^2} \frac{ \partial^2 {P}}{ \partial t ^2 } = 0$, 
where ${P}(x,y,z,t)$ represents the acoustic pressure scalar field in space and time, %the Laplacian $\nabla^2 {P}$ is taken with respect to the spatial coordinates,
and $c$ is the propagation speed in the medium.

When ultrasonic waves propagate through an absorbing medium, the initial pressure, $P_0$ reduces to $P(d)$ at a distance $d$  \cite{attref} as $P(d) = P_{0}e^{- \alpha d}$, where $\alpha$ ($\mathrm{ [Np \cdot cm^{-1}]}$) is the amplitude attenuation coefficient that captures energy dissipation from the ultrasonic beam and is a function of the carrier frequency $f$ as $\alpha = a f^{b}$ \cite{attref}, where $a$ ($\mathrm{ [ Np \cdot m^{-1}\ MHz^{-b}]} $) and $b$ are tissue attenuation parameters.

 \begin{table}[ht]
\caption{{\footnotesize Frequency Limits for A=100 $dB$}}
\centering
\begin{tabular}{c c c}
\hline\hline
Communication Range & Distance & Frequency Limit\\ [0.5ex] % inserts table %heading
\hline
Short Range & $\mu$m\ -\ mm & $>1GHz$ \\
Medium Range & mm\ -\ cm &  $\simeq100MHz$ \\
Long Range & $ > $cm & $\simeq10MHz$ \\[1ex]
\hline
\end{tabular}
\label{table_freq}
\end{table}

In \cite{wons} we observed that attenuation can be significant and it increases with the  distance between transmitter and receiver.
Moreover the beam spread is inversely proportional to the ratio between the diameter of the radiating surface and the wavelength of the operating frequency \cite{wons}.
Consequently, since most biomedical sensing applications require directional transducers, one needs to operate at high frequencies to keep the transducer size small. However, as expected, higher frequencies lead to higher attenuation.

In Table \ref{table_freq}, we summarize our findings in \cite{wons} on the maximum ``allowed'' carrier frequency for a $100 \:\mathrm{dB}$ maximum tolerable attenuation. 
For distances ranging between some $\mathrm{\mu m}$ up to a few $\mathrm{mm}$ (i.e., short range communications) frequencies higher than $1\:\mathrm{GHz}$ are allowed. When distances are higher than $1\:\mathrm{mm}$ but still lower than some $\mathrm{cm}$, i.e., for medium range communications, the transmission frequency should be decreased to approximately $100\:\mathrm{MHz}$. For distances higher than a few $\mathrm{cm}$, i.e., for long range communications, the transmission frequency should not exceed $10\:\mathrm{MHz}$. Our results in \cite{wons} are independently confirmed in \cite{Hogg}.

\section{Channel Modeling} \label{sec:channel}
To characterize the ultrasonic communication channel in tissues - for which, unfortunately, there is basically no literature available to date, except for \cite{Davilis} - we derive a deterministic channel model based on  wave propagation theory. 
Propagation  of acoustic waves through biological tissues is governed by three coupled first-order equations, i.e.,  the {\it continuity equation}, the {\it force equation} and the {\it equation of state} \cite{propagat}, which represent relationships among acoustic pressure, $P$, acoustic particle velocity $u$, and medium density $\rho$, and can be rearranged to obtain the Helmhotz equation in Section \ref{propagation}. The solution represents the acoustic field in time, evaluated at each spatial coordinate of the propagation medium. 		
\begin{figure*}[t!]
\centering
\subfigure[\label{mask_kwave}]
{\includegraphics[width=4.2cm, height=4.4cm]{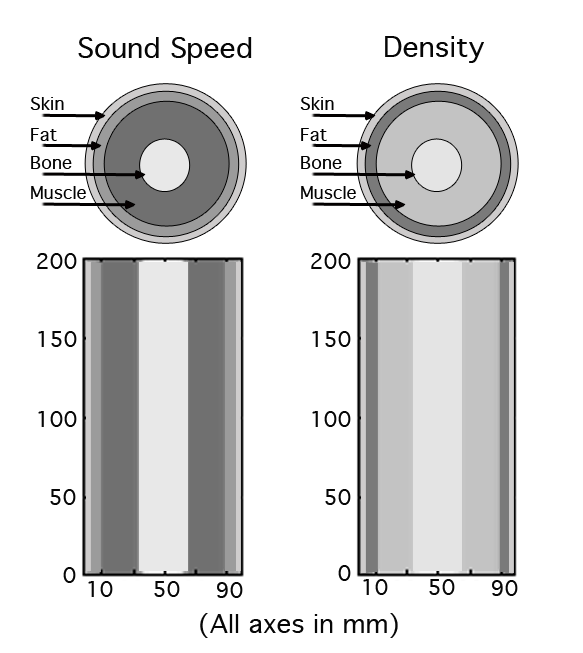}}
\hspace{0mm} %6
\subfigure[\label{kwave1}]
{\includegraphics[width=6.5cm]{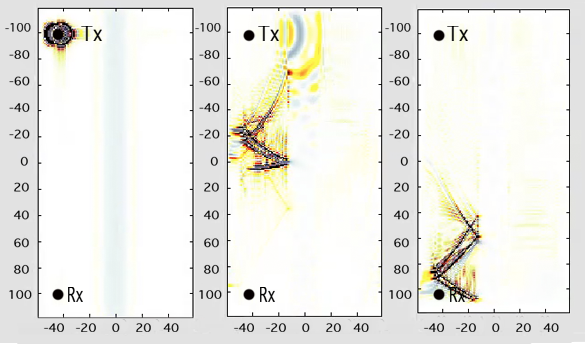}} %\hspace{0mm} %3
\subfigure[\label{impulse_resp}]
{\includegraphics[width=5.8cm]{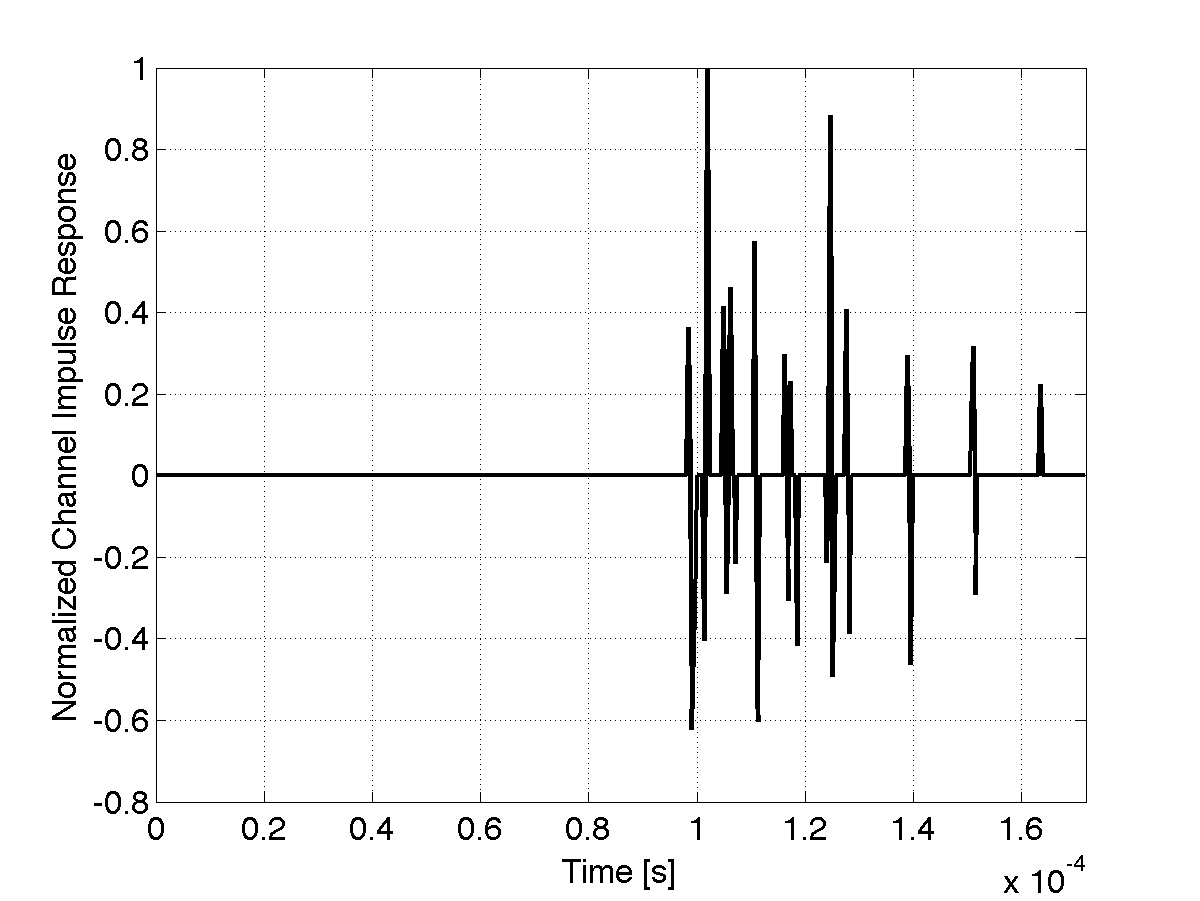}} %\hspace{0mm}
\caption{{\footnotesize \bf{ (a) Sound speed and density distributions
in a simulated arm. (b) Snapshots of the time evolution of acoustic
field. (c) Resulting ultrasonic channel impulse response in the
arm.}}}\vspace{-1mm}
\end{figure*}

Traditionally,  partial differential equations are solved using numerical methods such the finite-difference-method (FDM). 
We take a different, computationally more efficient approach based on the pseudo-spectral and k-space methods \cite{PS_kwave}.  Basically, the pseudo-spectral (PS) method reduces computational complexity in the spatial domain by using Fourier series expansions and FFTs, 
while the k-space method operates in the time domain by using k-space propagator functions (instead of classical finite differences) to approximate temporal derivatives. As a consequence, larger time steps can be used, reducing the simulation time with controllable accuracy. A powerful tool implementing this method is k-Wave \cite{kwave}.

We modeled a reference ultrasonic propagation channel in the human body to derive an accurate characterization of the channel impulse response, which was then used in PHY and MAC layer simulation studies. Specifically, we modeled a section of the human arm, including bones, muscles, fat and skin. We considered an heterogeneous  two-dimensional rectangular area of length $20\:\mathrm{cm}$ and width $10\:\mathrm{cm}$,  where density and sound velocity are distributed as shown in Fig. \ref{mask_kwave}.  The bone half-section has a $18\:\mathrm{mm}$ width, the muscle half-section has a $22\:\mathrm{mm}$ width, while fat and skin have a half-section of $7\:\mathrm{mm}$ and $3\:\mathrm{mm}$ width, respectively.  The medium parameters, sound speed $c$, density $\rho$ and attenuation $a$ and $b$ are as in \cite{atable}. 

We obtained the channel impulse response by releasing an ideal Dirac pulse in the left top corner of the muscle section. The receiver is located in the left bottom corner of the muscle section. Transmitter and receiver are located $20\:\mathrm{cm}$ away from one another. In Fig. \ref{kwave1}, we show snapshots of the acoustic field time propagation in the modeled environment. We observe that the effect of the bone and tissues is to partially reflect and scatter the acoustic wave transmitted by the source.  
We then obtain the channel impulse response by recording the time series of the signal at the receiver sensor. 

Figure \ref{impulse_resp} reports the resulting impulse response in this scenario. We observe that the effect of multipath and scattering is to introduce attenuated signal replicas spaced in time. Because of the very short duration of the transmitted pulses, replicas do not interact destructively, which provides a strong motivation for the proposed transmission scheme presented in Section \ref{mac}.  We can model the channel response as a complex-valued low-pass equivalent impulse response. 
We can then characterize the channel through the mean excess delay ($\tau_m$) and the RMS delay spread ($\tau_{RMS}$). For the channel simulated above, we obtained $\tau_m = 1.2779 \cdot 10^{-5}s$ and $\tau_{RMS} = 2.6883\cdot 10^{-5}s$. Since the coherence bandwidth of the channel is proportional to the inverse of $\tau_{RMS}$, we should consider the above channel as frequency selective for signals of bandwidth above approximately 7 kHz.

\section{Ultrasonic Wideband} \label{mac}
Based on the considerations summarized in Sections \ref{propagation} and \ref{sec:channel}, we design and propose a new ultrasonic transmission and multiple access technique, which we refer to as Ultrasonic WideBand (UsWB). The key design objectives of UsWB are
(i) to enable low-complexity and reliable communications in ultrasonic channels against the effect of multipath reflections within the human body;  (ii) to limit the thermal effect of communications, which is detrimental to human health; (iii) to enable distributed medium access control and rate adaptation to combat the effect of interference from co-located and simultaneously transmitting devices.  
UsWB is jointly designed to provide physical layer functionalities and medium access control arbitration and adaptation to enable multiple concurrent co-located transmissions with minimal coordination. 

{\bf Ultrasonic Pulsed Transmissions. } Ultrasonic wideband is based on the idea of transmitting very short ultrasonic pulses following an {\it adaptive} time-hopping pattern together with a superimposed adaptive spreading code. 
Baseband pulsed transmissions enable high data rate, low-power communications, low-cost transceivers, and have been proposed for RF short-range, high data rate communications \cite{uwbmain, cuomo}, although with much shorter pulse durations (and consequently larger bandwidth) than achievable in ultrasonic communications. 

The characteristics of pulsed transmissions appear to ideally address the requirements discussed above. Their fine delay-resolution properties are well-suited for propagation in the human body, where inhomogeneity in terms of density and propagation speed, as well as the pervasive presence 
of very small organs and particles, cause dense multipath and scattering. 
When replicas of pulses reflected or scattered are received with a differential delay at least equal to the pulse width, they do not overlap in time with the original pulse. Therefore, for pulse durations in the order of hundreds of nanoseconds \cite{transd}, pulse overlaps in time are reduced and multiple propagation paths can be efficiently resolved and combined at the receiver to reduce the bit error rate.  Also, the low duty cycle of pulsed transmissions reduces the impact of thermal and mechanical effects, which are detrimental for human health \cite{duty_temp}.  
Finally, carefully designed interference mitigation techniques may enable MAC protocols that do not require
mutual temporal exclusion between different transmitters. This is crucial in the ultrasonic transmission medium since the propagation delay is five orders of magnitude higher than in RF in-air channels and carrier-sense-based medium access control protocols are ineffective \cite{BasagniBP04}.  In addition, data rate can be flexibly traded for power spectral density and multipath performance.

\subsection{Physical Layer Model} \label{uswb}

{\bf Adaptive Time-Hopping. } Consider a slotted time divided in chips of duration $T_c$,  with chips 
organized in frames of duration $T_f=N_h \cdot T_c$, where $N_h$ is the
number of chips per frame. % If all users have the same frame length,
 Each user transmits one pulse in one chip
per frame, and determines in which chip to transmit based on a
pseudo-random {\it time hopping sequence} (THS), i.e., a sequence generated by seeding a random number generator with the user's unique ID. 

The train of pulses is modulated
based on pulse position modulation (PPM), i.e., a `1' symbol is
carried by a pulse delayed by a time $\delta$ with respect to
the beginning of the chip, while a `0' symbol begins with the chip.
The signal $s^{(k)}(t,i)$ generated by the
$k^{th}$ user to convey the $i^{th}$ symbol is expressed as
\begin{equation}
\label{signal_new}
s^{(k)}(t,i) =p(t-c^{(k)}_iT_c-iT_f-d^{(k)}_i\delta),
\end{equation}
where $p(t)$ 
is the second derivative of a Gaussian Pulse, 
$\{ c_i^{(k)}\}$ is the
time hopping sequence of the $k^{th}$ source, with $0 \leq c_i^{(k)}
\leq N_h-1$, and $\{d^{(k)}_i\}$ is the information-bearing sequence,
$d^{(k)}_i \in \{0,1\}$. 
The resulting data rate, in pulses per second, is expressed as $R(N_h)=\frac{1}{T_f} = \frac{1}{N_hT_c}$.
By regulating the TH frame length $N_h$, i.e., the average inter-pulse time, a user can adapt its transmission rate, and as a consequence modify the average radiated power and therefore the level of interference generated to other ongoing communications. We observe at this point that an individual user has little incentive to increase its frame size, since that results in a lower achievable data rate, without any major benefit for the user itself (since the level of interference perceived depends primarily on the frame length of the other users, and not on its own). However, a longer time frame reduces the interference generated to the other users. Therefore, selfish/greedy frame adaptation strategies do not work well in this context - cooperative strategies are needed. 

{\bf Adaptive Channel Coding. } An adaptive channel code \cite{dcc} can further reduce the effect of mutual interference from co-located devices, by dynamically regulating the coding rate to adapt to channel conditions and  interference level. 
Various channel coding solutions have been proposed \cite {uwbmain, dcc} with different performance levels and computational complexity. We rely on simple {\em pseudo-orthogonal spreading codes} because of their excellent multiple access performance, limited computational complexity, and inherent resilience to multipath.  Each symbol (i.e., bit) is spread by multiplying it by a pseudorandom code before transmission. At the receiver side, with prior knowledge of the code used at the transmitter, the signal can be de-spread, and the original information recovered. With different orthogonal (or pseudo-orthogonal) codes, multiple nodes can transmit simultaneously on the same portion of the spectrum, with reduced interference.
We explored two alternative modulation schemes, which we refer to as PPM-BPSK-spread and PPM-PPM-spread.

In PPM-BPSK-spread, the binary spreading code at the $k^{th}$ node, $\{a^{(k)}_{j}\}$, is defined as a pseudorandom code of  $N_s$ chips with $a_{j}\in\{-1,1\}$.  Accordingly, the information bit is spread using BPSK modulated chips, and by combining with  time hopping, (\ref{signal_new}) can be rewritten as
\begin{equation}
\label{signal_ds}
s^{(k)}(t,i) = \sum_{j=0}^{N_s-1} a^{(k)}_{j} {p(t-c^{(k)}_jT_c-jT_f-d^{(k)}_i\delta)},
\end{equation}
where $\delta$ is the PPM displacement of a pulse representing a `1' bit, while chip information is carried in the pulse polarity.

In PPM-PPM-spread, the information bit is spread using PPM-modulated chips. In this case, the binary spreading code can be defined as a pseudorandom code of  $N_s$ chips with $a_{j}\in\{0,1\}$. 
With time hopping, (\ref{signal_new}) can be rewritten as
\begin{equation}
\label{signal_ds}
s^{(k)}(t,i) = \sum_{j=0}^{N_s-1}  {p(t-c^{(k)}_jT_c-jT_f-(a^{(k)}_{j} \oplus d^{(k)}_i)\delta)}
\end{equation}
where $ \oplus$ represents a $XOR$ operation.

\begin{figure}
\centering
\includegraphics[width=3.5in]{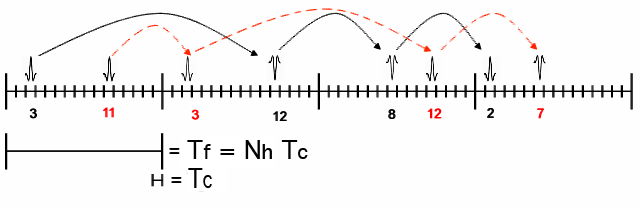}
\caption{{\footnotesize \bf{ Example of two ongoing transmissions using time hopping sequences $TH_1=\{3, 12, 8, 2\}$ and $TH_2=\{11,3, 12, 7\}$ and PPM-BPSK spreading codes $SC_1=\{ 1, -1, -1, 1\}$ and $SC_2=\{ 1, 1, 1, -1\}$.}}}
\label{th-ds}
\end{figure}

In Fig. \ref{th-ds} we show an example of a combined time hopping and PPM-BPSK-spread coding strategy.
Since the spreading operation associates $N_s$ chips to one information bit, the information rate will be further reduced by a factor $N_s$, i.e., 

\begin{equation}
\label{rate_code}
R(N_h,N_s) = 	\frac{1}{N_sT_f}=\frac{1}{N_sN_hT_c},
\end{equation}
while the energy per bit is increased by a factor $N_s$. Note that there is  a tradeoff between robustness to multi-user interference (which increases with longer spreading codes), and energy consumption and information rate. In Section \ref{adapt} we discuss joint dynamic adaptation of  frame and code length. 
With BPSK-modulated chips the absolute received {\em phase information} is needed for decoding. Therefore, a {\em coherent} receiver with accurate channel knowledge is needed. Instead, with a pure PPM-modulated signal, a simple {\em non-coherent} energy detector receiver is sufficient. . 
The latter requires frame synchronization only, and its hardware complexity is significantly lower \cite{uwb_hw}.

{\bf Signal to Interference-plus-Noise Ratio.} 
We can express the signal to interference-plus-noise ratio (SINR) for impulsive transmissions at the receiver of link $i$ as \cite{uwbmain}
\begin{equation} \label{snr_new}
SINR_i({\bf N_{h},N_{s}}) = N_{s,i} \frac{P_ig_{i,i}\ N_{h,i}T_c}{\eta+\sigma^2 T_c\sum_{k\in  \mathcal{I}_i} \frac{N_{h,i}}{N_{h,k}} P_kg_{k,i}},
\end{equation}
where $P_i$ is the average power per pulse period emitted by the $i^{th}$ transmitter, $g_{i,j}$ is the path gain between the $i^{th}$ transmitter to the $j^{th}$ receiver, $\eta$ represents background noise energy and $\sigma^2$ is an adimensional parameter depending on the shape of the transmitted pulse and the receiver correlator. The set $\mathcal{I}_i$ represents the set of links whose transmitter interferers with the receiver of link $i$. Remind that $T_{f}=N_hT_c$ is the frame duration. 
The expression in \eqref{snr_new} depends on the vectors of frame and code lengths of all the ongoing communication in the network, i.e., ${\bf N_{h},N_{s}}$.
The term $\frac{N_{h,i}}{N_{h,k}}$ accounts for the level of interference generated by each interferer $k$ to the receiver of link $i$, i.e., the number of pulses transmitted by the $k^{th}$ transmitter during the time frame of the $i^{th}$ user. Note that we will not consider power control strategies in our treatment, since \cite{cuomo} showed that in the linear regime, when the objective is to maximize the aggregate data rate, the optimal solution always corresponds to points where individual devices transmit at the maximum power, or do not transmit at all. 

Note that increasing (or decreasing) the spreading code of the node of interest, $N_{s,i}$, leads to an increase (decrease) in the SINR. When the node of interest increases (decreases) its frame length, $N_{h,i}$, while the other nodes do not, we expect no variation in the SINR (there is in fact a slight increase (decrease) in the SINR, which can be neglected under high SNR conditions, $\eta \ll \sum_{k\in  \mathcal{I}_i} P_kg_{ki}$). Finally, when the frame length of the interfering nodes is increased (decreased), the SINR increases (decreases).

\section{MAC and Rate Adaptation}
\label{adapt}

In this section, we discuss UsWB medium access control principles and rate adaptation. Based on our discussion so far, there is a tradeoff between (i) resilience to interference and channel errors, (ii) achievable information rate, and (iii) energy efficiency. In this section, we introduce medium access control and rate adaptation strategies designed to find optimal operating points along efficiency-reliability tradeoffs. We first consider rate-maximizing adaptation strategies in Section \ref{sec:mac:rate}. Then, we propose two different energy-minimizing strategies in Section \ref{sec:mac:energy}.

\subsection{Distributed Rate-Maximizing Adaptation} \label{sec:mac:rate}
The objective of the rate-adaptation algorithm under consideration is to let each active communication maximize its transmission rate by selecting a pair of code and frame lengths, based on the current level of interference and channel quality measured at the receiver and on the level of interference to the other ongoing communications generated by the transmitter.   
We consider a decentralized ultrasonic intra-body area network. Denote by $\mathcal{N}$  the set of $|{\mathcal{N}}|$  existing connections, and by $N_{h, max}$ and $N_{s, max}$ the maximum lengths supported,
\begin{equation}\label{domain_const1}
0<N_{h,i}\le N_{h, max},\ \forall i \in  \mathcal{N},\ N_h \in \mathbb{N},
\end{equation}
\begin{equation}\label{domain_const2}
0<N_{s,i}\le N_{s, max},\ \forall i \in  \mathcal{N},\ N_s \in \mathbb{N},
\end{equation}
where $\mathbb{N}$ is the set of natural numbers. According to the transmission scheme discussed in Section \ref{uswb}, each node transmits at a rate $R$  as in (\ref{rate_code}) and each receiver experiences an SINR  as in (\ref{snr_new}). Each node has a minimum data rate requirement, i.e., $R_i(N_{h,i}, N_{s,i})\ge R_{min}$, and a minimum SINR requirement, i.e., $SINR_i({\bf N_{h,i}, N_{s,i}})\ge SINR_{min}$. %$SINR(N_{h,i}, N_{s,i})\ge SINR_{min}$ .

{\bf Explicitly Cooperative Problem.} Denote the frame length and the code length selected by the receiver of the communication $r$ as $N_{h,r}$ and $N_{s,r}$. 

The objective of each user is to locally optimize the information rate of the communication by solving the following problem: 
\begin{align}%\nonumber
\underset{{N_{h,r}, N_{s,r}}}{\text{maximize}}\ & R_r (N_{h,r},N_{s,r}) \label{distr_prob} \\
\text{subject \ to}\ &  R_r(N_{h,r}, N_{s,r}) \ge R_{min} \label{rmin} \\
&  SINR_r(N_{h,r}, N_{s,r}) \ge SINR_{min} \label{sinr_const_game_r} \\
&  SINR_i(N_{h,r}, N_{s,r}) \ge SINR_{min}\  \ \forall i \in  \mathcal{I}_r \label{sinr_const_game_i} 
\end{align}
where $\mathcal{I}_r$ is the set of the connections interfering with the $r^{th}$ connection. The constraints on the maximum frame and code length in \eqref{domain_const1} and \eqref{domain_const2} are also implicitly considered. We refer to this as the {\em explicitly cooperative problem}.

{\bf Implicitly Cooperative Problem.} If all nodes measure the same level of interference, that is, all network nodes are close enough to be all in the same transmission range, and all nodes have the same minimum rate and minimum SINR requirements,  the level of interference that can be tolerated by each receiver is the same. Therefore, this information does not need to be exchanged. The problem in \eqref{distr_prob} becomes
\begin{eqnarray}\nonumber \label{distr_prob2}
&\underset{{N_{h,r}, N_{s,r}}}{\text{maximize}}\ & R_r (N_{h,r},N_{s,r})\label{max_game2} \\
&\text{subject\ to} & R_r(N_{h,r}, N_{s,r})\ge R_{min}\\ \label{rate_const2}
& & SINR_r(N_{h,r}, N_{s,r}) \ge SINR_{min} \label{sinr_const_game2} 
\label{eq:fmlt sub21}
\end{eqnarray}
The system of SINR inequality constraints in \eqref{distr_prob} becomes here a single inequality constraint. The new problem can be interpreted as {\it finding the optimal pair of code and frame length that maximize the rate, given a minimum rate and a minimum SINR, under the assumption that all the other nodes will be acting in the same way}. As we will show in Section \ref{performanceevaluation}, since the problem to be solved is the same for each node, a globally optimal pair of code and frame lengths will be found. We refer to this as the {\em implicitly cooperative problem}.

We now provide a detailed analytical description of the explicitly cooperative problem. The implicitly cooperative problem can then be derived as a special case.

The explicitly cooperative problem as stated in \eqref{distr_prob} is an integer program, i.e., variables $N_h$ and $N_s$ are integer. If the domain of the problem is small, 
 it can be solved by enumeration, i.e., trying all the possible combinations of $N_h$ and $N_s$. When the domain of the problem increases in size 
 a relaxation method can be used to transform the integer problem into a fractional problem. The relaxation operation consists of replacing the constraints in \eqref{domain_const1} and (\ref{domain_const2}) with  
\begin{equation}
0<N_{h}\le N_{h, max},\  0<N_{s}\le N_{s, max},\  \ N_s,N_h \in \mathbb{R}\label{ns_game_real},
\end{equation}

where $\mathbb{R}$ is the set of real numbers. Note that the optimal solution to the relaxed problem is not necessarily integral. However, since the feasible
set of the relaxed problem is larger than the feasible set of the original integer program, the optimal value of the former, $p^*_{rlx}$ is 
a lower bound on the optimal value of the latter, $p^*_{int}$, i.e., $L = p^*_{rlx} \le p^*_{int}$.
The relaxed solution can be used to guess the integer solution by rounding its entries based on a threshold $t\in[0,1]$. %, i.e.,

If the rounded solution, $\hat{x}^*_{int }$, is feasible for the original problem, i.e., all the constraints are satisfied, then it can be considered a guess at a good, if not optimal, point for the original problem. Moreover, the objective function evaluated at $\hat{p}^*_{int }$ is an upper bound on $p^*_{int}$, $U = \hat{p}^*_{int} \ge p^*_{int}$
Note that $\hat{x}^*_{int }$ is cannot be more than (U-L)-suboptimal for the original problem.
We can express the relaxed problem as a geometric program  \cite{geometric}, that is, minimizing a posynomial function under posynomial inequality constraints. 
First,  the problem must be restated in terms of minimizing the inverse of the data rate, that is, $R^{-1}_r(N_{h,r},N_{s,r})$. Then, the new objective function and all the constraints can be expressed in monomial and posynomial form. The objective function, as well as the minimum rate function, is a monomial function, $N^{-1}_{h,r}\ N^{-1}_{s,r}$. The same holds for the constraint functions defined is \eqref{ns_game_real}, and for the SINR functions defined in \eqref{snr_new}.

Let us first consider the SINR constraint in \eqref{sinr_const_game_r}, and note that in the distributed optimization problem the only variables are the code and the frame length of the receiver of interest, $N_{s,r}$ and $N_{h,r}$. If we define 
\begin{equation*}
\alpha_r = P_r g_{r,r} T_c, \ \ \ \beta_k = {\sigma^2T_c P_{k}g_{r,k}}/ {N_{h,k}},
\end{equation*}
then \eqref{sinr_const_game_r} can be rewritten as 
a posynomial constraint
\begin{equation} \label{ber}
\eta N_{s,r}^{-1}N_{h,r}^{-1} +N_{s,r}^{-1}\sum_{k\in  \mathcal{I}_r} \beta_k \le \frac{\alpha_r}{SINR_{min}}.
\end{equation}

For the remaining ($|{\mathcal{I}_r }|-1$) SINR constraints, if we define 
\begin{eqnarray}\nonumber \label{distr_prob2}
&\gamma_i =  P_i g_{i,i} T_c N_{h,i} N_{s,i}\ \ \ \epsilon_i = {\sigma^2T_c P_{r}g_{i,r}} {N_{h,i}}\\
&  \delta_i = \sum_{k\in  \mathcal{I}_i}  \sigma^2T_c P_{k}g_{i,k} {N_{h,i}}/{N_{h,k}}
\label{eq:fmlt sub21}
\end{eqnarray}
then \eqref{sinr_const_game_i} %for  $i\in \mathcal{I}_r$ 
can be rewritten as
\begin{equation}
\frac{\eta+ \delta_i +  \epsilon_iN^{-1}_{h,r}}{\gamma_i} \le  \frac{1}{SINR_{min}},
\end{equation}
which is a linear, and thus monomial, constraint. In particular, this constraint shows that the frame length variation only leads to an increase (decrease) in the level of interference produced. Moreover, the constraint suggests that the optimal frame length is upper bounded by the co-located node that is experiencing the highest level of interference.
Based on these observations, the cooperative optimization problem becomes
\begin{align}%\nonumber \label{distr_prob4}
\underset{{N_{h,r}, N_{s,r}}}{\text{minimize}}\ & R^{-1}_r(N_{h,r},N_{s,r})\label{max_game4} \\ 
\text{subject\ to}\  & R^{-1}_r(N_{h,r}, N_{s,r})\le R^{-1}_{min}\\
 & \eta N_{s,r}^{-1}N_{h,r}^{-1} +N_{s,r}^{-1}\sum_{k\in  \mathcal{I}_r} \beta_k  \le \frac{\alpha_r}{SINR_{min}}   \label{sinr_const_game4_1} \\
 & N_{h,r}  \ge \frac{\epsilon_i} {\gamma_i\ SINR^{-1}_{min}-\eta -\delta_i} \label{sinr_const_game4_2}.
\end{align}

Finally, a geometric program can be in general transformed into a convex program through a logarithmic transformation of the optimization variables, which can be solved in polynomial time through interior-point methods \cite{interior}.
\subsection{Distributed Energy-minimizing Rate Adaptation}\label{sec:mac:energy}

We now concentrate on rate adaptation with the objective of reducing the energy consumption of  UsWB. 
We define (i) $E_p$, the energy per pulse; (ii) $E_b$, the energy per bit, i.e., $E_b =E_p \cdot  N_s$; and (iii) $E_s$, the average energy emitted per second, i.e., $E_s =  E_p/ (T_c \cdot N_h)$.

The energy per bit is a linear function of the spreading code length, therefore of the number of pulses transmitted per each bit. The average power emitted per second is a function of the inverse of the frame length, hence of the number of pulses transmitted per second. However, both depend on the value of $E_p$, the energy transmitted per pulse, which is related to the (electrical) power absorption of the ultrasonic transducer.

{\bf Modeling the Energy Consumption of a Piezoelectric Transducer. } 
We consider electronically-driven piezoelectric transducers. Since their electronics are known to have minimal current leakage, the majority of power consumption comes from the piezoelectric element. %An accurate analysis of the energy cons
Outside the region of resonance, a piezoelectric ceramic transducer can be viewed (from the electrical point of view) as a parallel plate capacitor with capacitance $C_0$. Thus, the main source of power consumption comes from charging such capacitor \cite{power}. 
Then, ignoring charge and discharge losses, and considering a capacitor with voltage supply $V$ and pulse repetition frequency $f$ (which corresponds to the charge and discharge frequency of the capacitor), the power consumption  $P_c$ can be expressed as
\begin{equation}
\label{power_eq}
P_c ={f\hspace{0.3mm}C_0\hspace{0.3mm}V^2}.
\end{equation} 
The static capacitance and the voltage supply values should be based on transducer-specific considerations. The static capacitance value of a disc-shaped transducer is given as 
$C_0 = \frac{A\hspace{0.3mm} \epsilon_0\hspace{0.3mm} K}{t_h}  \hspace{3mm} [F]$, where $\epsilon_0$ is the permittivity in free air ($8.8542\:\mathrm F/m$), $K$ is the dielectric constant of the material (adimensional), $A$ [$m^2$] is the area of the disc, and $t_h$ [$m$] is its thickness. 
The dielectric constant $K$ depends on  material, frequency and mechanical state of the transducer.

We derive appropriate limits for the voltage supply based on safety concerns, which impose a limit on the radiated acoustic power. As reported in \cite{Hogg}, no tissue damage occurs in intra-body ultrasonic propagation as long as the acoustic power dissipation in tissue is limited to $10^4\:\mathrm{W/m^2}$. From this limit, we derive the corresponding maximum pressure magnitude that can be radiated by the transducer, and consequently the maximum voltage input.
For example, if we consider arm muscle as the primary propagation medium, through the quadratic relation between the acoustic intensity and the acoustic pressure, 
$I = {(P_{RMS})^2}/{\rho c}$, along with density and speed of sound parameters reported in \cite{atable}, the maximum pressure magnitude that can be radiated by the transducer is found to be approximately $0.13\:\mathrm{MPa}$.

We can derive the related transducer voltage input, corresponding to the maximum radiated pressure, through the constitutive equation of piezoelectric materials.  
The latter expresses the relationship among  mechanical strain and electrical displacement for the piezoelectric element considering the electrical and mechanical stress  and is usually expressed in tensor notation \cite{piezo_tensor}. Here, we are interested in the so-called {\it converse effect}, i.e.,  the material strain proportional to an applied voltage. If we consider a symmetric crystal structure, the constitutive relations reduce to a few parameters.  %$d_{33}$ ( in $[C/m]$) and
In particular, the so-called $g_{33}$ parameter $[Vm/N]$ represents the piezoelectric voltage coefficient, when the polarization field and the piezoelectrically induced strain are both parallel to the disc axis (usually referred to as the third axis). Based on this, one can express the electric field along the third axis $E \hspace{1mm} \mathrm{[V/m]}$ as $E = g_{33}\hspace{0.3mm}P$,
where $P$ again represents the output pressure of the transducer. Since the electric field $E$ can in turn be expressed as the ratio between voltage and electrode distance, we have $V = g_{33}\hspace{0.3mm}P\hspace{0.3mm}t_h$.
Finally, from (\ref{power_eq}), the energy consumed to generate a single pulse is obtained as
\begin{equation}
E_p = C_0  \hspace{1mm} (g_{33}\hspace{0.3mm}P\hspace{0.3mm}t_h)^2  \hspace{3mm} [J].
\end{equation}

{\bf Energy-minimizing Rate Adaptation}. Based on this model, we design a rate adaptation strategy where the objective of $r$ is to minimize (i) the energy per bit, $E_b$, or (ii) the average energy emitted per second $E_s$. The problem consists of finding the optimal frame length $N_h$ and the optimal spreading code length $N_s$ that minimize $E_b$ ($E_s$) while meeting the minimum SINR constraints and keeping the data rate over a given threshold. The energy problem is formally expressed below.
\begin{align}\nonumber
\underset{{N_{h,r}, N_{s,r}}}{\text{minimize}}\  & E_b(N_{s,r})\ (\mathrm{or}\ E_s(N_{h,r}))  \label{energyb_game} \nonumber \\
\text{subject\ to}\ & \eqref{rmin}\ \eqref{sinr_const_game_r}\ \eqref{sinr_const_game_i} \nonumber
\end{align}

Similarly, the problem above can be relaxed to a geometric program as discussed in Section \ref{sec:mac:rate}. 

\subsection{Medium Access Control Protocol} \label{protocol}
In UsWB, distributed medium access control coordination is achieved by exchanging information on logical {\it control channels}, while data packets are transmitted over logical {\it data channels}. 
We consider unicast transmissions between a transmitter $Tx$ and a receiver $Rx$. 
When $Tx$ needs to transmit a packet, it first needs to reserve a dedicated channel to $Rx$. The connection is opened through the common control channel, which is implemented through a unique TH-sequence and a spreading code known and shared by all network devices.

In the two-way handshake procedure, $T_x$ sends a {\tt Request-to-Transmit (R2T)} packet to $R_x$, which contains its own ID. If $R_x$ is idle, a {\tt Clear-to-Transmit  (C2T)} control packet is sent back to $T_x$. In case of failure and, thus, timer expiration, $T_x$ will attempt a new transmission after a random backoff time, for a maximum of $N_R$ times. During these initial steps, since an estimate of the current interference is not available, the transmitter transmits at the minimum data rate, conservatively using a maximum frame length and spreading code length,  thus generating low interference. 
After receiving the {\tt C2T}, the transmitter switches to a dedicated channel by computing its own TH sequence and spreading code obtained by seeding a pseudo random sequence generator with its ID.   
As a consequence, both $T_x$ and $R_x$  leave the common channel and switch to a dedicated channel.  
Once the connection has been established, $T_x$ sends the first packet using maximum frame  and spreading code length. The receiver $R_x$ computes the optimal frame and spreading code lengths as discussed in Section \ref{uswb}. This information is piggybacked into {\tt ACK} or {\tt NACK} packets. 

In the explicitly cooperative case, discussed in Section \ref{sec:mac:rate}, once the communication has been established, $R_x$ does not leave the common control channel. Instead, it keeps ``listening" to both the dedicated and common control channel at the same time. In the dedicated control channel, $R_x$ sends to $T_x$ the optimal frame and code lengths to be used for the next transmission. In the common control channel, $R_x$ exchanges with other co-located receivers information on the level of tolerable interference. 
%%%%%%%%%%%%%%%%

%\section{Performance Evaluation} \label{performance}
\section{Performance Evaluation} \label{performanceevaluation}

In this section we evaluate the proposed system through a custom-designed multi-scale simulator that considers UsWB performance at three different levels, i.e., (i) at the {\it  acoustic wave level} by modeling ultrasonic propagation in tissues through reflectors and scatterers, (ii) at the {\it bit level} by simulating in detail the  ultrasonic transmission scheme, (iii) at the { \it packet level} by simulating networked operations and distributed medium access control and adaptation. 

\begin{figure}
\centering
\includegraphics[width=3.5in]{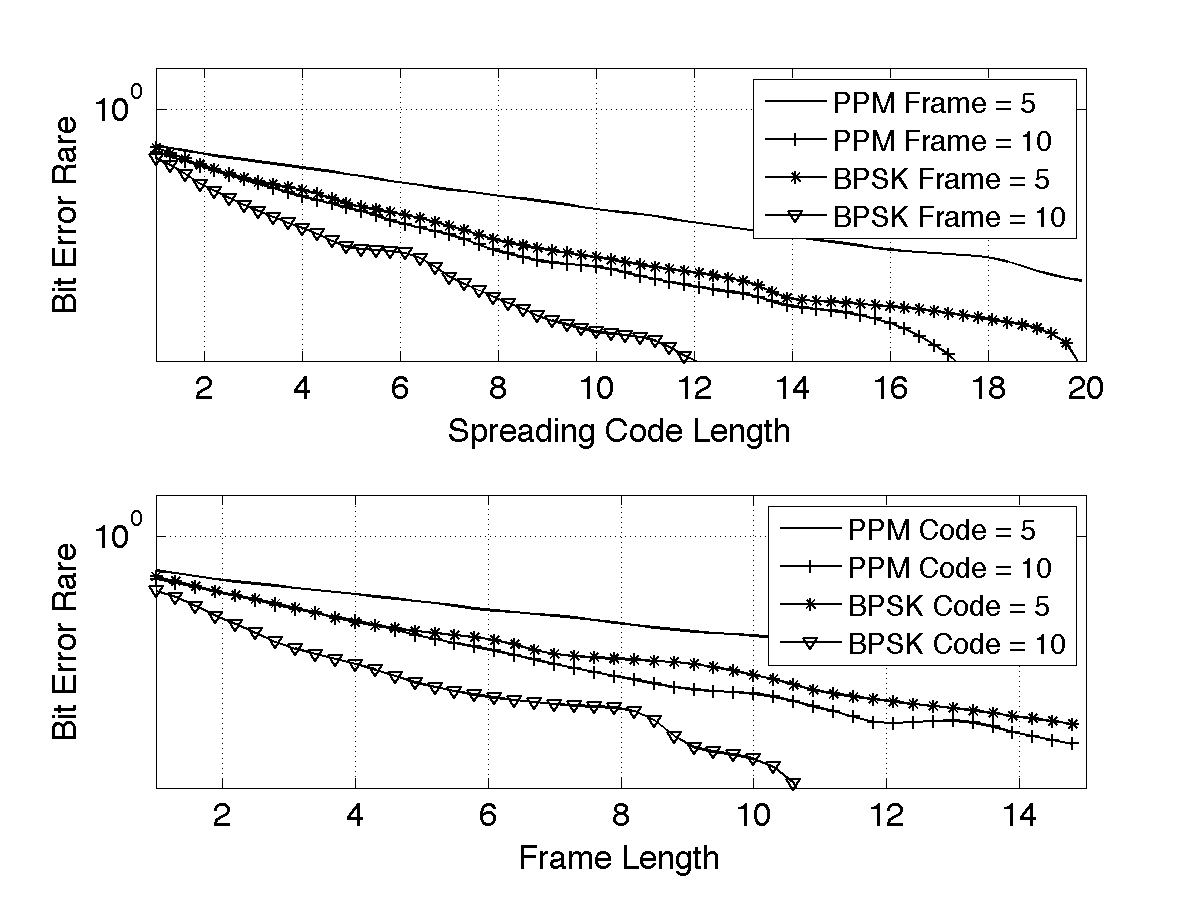}
\caption{ {\footnotesize {\bf (a) BER vs. TH frame length and (b) BER vs. spreading code length for PPM-BPSK spread and PPM-PPM spread.}}}
\label{ber_frame_code_4}
\end{figure}

The acoustic wave level simulation is performed as described in Section \ref{sec:channel}. A channel impulse response is obtained by simulating propagation in the human arm. Transmission at the bit level is modeled through a custom physical layer simulator, which allows to obtain an empirical model of the Bit Error Rate (BER) against different values of the TH frame length and spreading code length, for different levels of interference.  
The physical layer simulation models a transmitter and a receiver (located $20\:\mathrm{cm}$ apart) communicating over an acoustic channel through UsWB. Each sender node transmits information as discussed in Section \ref{mac}.  Simulations are performed to obtain an estimate of the achievable BER upon varying the frame length and the spreading code length with  a different number of interferers transmitting on the same channel. We implement both solutions presented in Section \ref{uswb}, i.e.,  PPM-BPSK-spread with coherent receiver and the PPM-PPM-spread with non-coherent receiver. An extensive simulation campaign was conducted and BER values were obtained as a function of the spreading code and frame lengths for different number of simultaneously active connections. Figure \ref{ber_frame_code_4} reports the BER values when 4 pairs of nodes are transmitting at the same time, and as expected the coherent PPM-BPSK-spread outperforms the non-coherent PPM-PPM-spread. 
The empirical model of the BER as a function of frame length and spreading code length for a given
level of interference  is then imported in a Java-based event-driven packet-level simulator, which models all aspects of the UsWB communication protocol stack.

{\bf Network Simulation Topology.} We considered two different settings for the network level simulations. First, we consider a 2-D topology with 18 static nodes randomly located inside a square of side $20\:\mathrm{cm}$. A maximum of nine communicating pairs are considered. We also assume that the transmission range is greater than the maximum distance between the nodes, thus all nodes measure the same number of interferers. 
The second setting consists of three 2-D squared clusters of nodes with side $10\:\mathrm{cm}$, displaced $20\:\mathrm{cm}$ apart from each other (distance center-to-center). In each cluster, the nodes are randomly deployed according to a gaussian distribution. The intermediate located cluster (\#1) contains one communicating pair, while the other clusters, (\#2) and (\#3), contain a variable number of communicating pairs. We also assume the transmission range to be of $30\:\mathrm{cm}$, thus all the nodes of adjacent clusters interfere with each other, while nodes of non adjacent clusters do not. Under this assumption, communicating pairs from different clusters measure different levels of interference.

During the simulations, the frame length and spreading code length are adapted to the interference level measured at the receiver according to one of the objectives discussed in Section \ref{sec:mac:rate}. In particular, the first setting matches the condition required by the implicitly cooperative problem in (\ref{distr_prob2}), while for the second setting the explicitly cooperative problem in (\ref{distr_prob}) is considered. 
The simulation time is set to $100\:\mathrm{s}$ and nodes start establishing connections at a random time instant but no later than $2\:\mathrm{s}$ after the simulation start time. We consider an infinite arrival rate at each transmitter, i.e., transmitters are always backlogged. The maximum allowed frame and spreading code length is set to 15 slots and 20 chips, respectively. The maximum supported rate, achieved when frame and spreading code length are both set to one, is equal to $2\:\mathrm{Mbit/s}$. 
The minimum SINR constraint is equivalent to a maximum BER constraint set to $10^{-6}$. 

\subsection{Rate Maximization}
\begin{figure}
\centering
\includegraphics[width=3.5in]{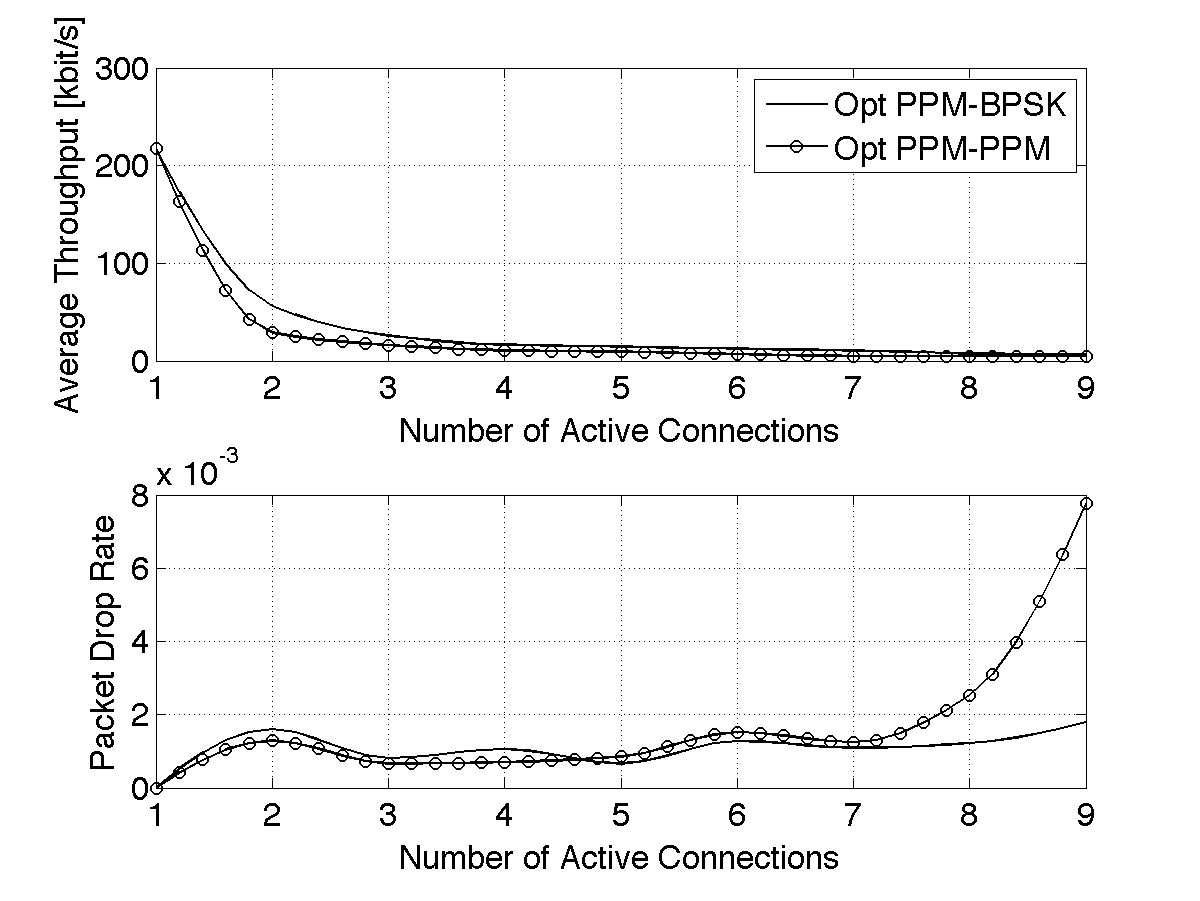}
\caption{ {\footnotesize {\bf   Throughput and packet drop rate with rate-optimal solution for implicitly cooperative problem.}}}
\label{thrgpt_drop_opt_ppmbpsk}
\end{figure}

{\bf Implicitly cooperative solution.} In Fig.  \ref{thrgpt_drop_opt_ppmbpsk} we plot and compare network throughput and packet drop rate for the rate-maximizing strategy, when frame and code length are adaptively regulated based on the implicitly cooperative problem in Section \ref{sec:mac:rate}. The rate-maximizing solution is presented for both the transmission schemes discussed in Section \ref{uswb}, the PPM-BPSK-spread with coherent receiver and the PPM-PPM-spread with non-coherent receiver. As expected the coherent PPM-BPSK solution performs better in terms of throughput. This happens because the BER constraint is satisfied for lower values of frame length and code length, which leads to higher data rates according to (\ref{rate_code}). 

In terms of packet drop rate, for the coherent PPM-BPSK system the BER constraint is satisfied for any number of active connections considered. Instead, for the non-coherent PPM-PPM system, when the number of active connections is greater than seven the BER constraint cannot be satisfied anymore, and therefore the packet drop rate increases.
Note that this problem can be overcome by simply relaxing the constraint on the maximum size of code length or frame length, without relaxing the constraint on the maximum BER. Clearly this leads to a lower data rate. 

\begin{figure}
\centering
\includegraphics[width=3.5in]{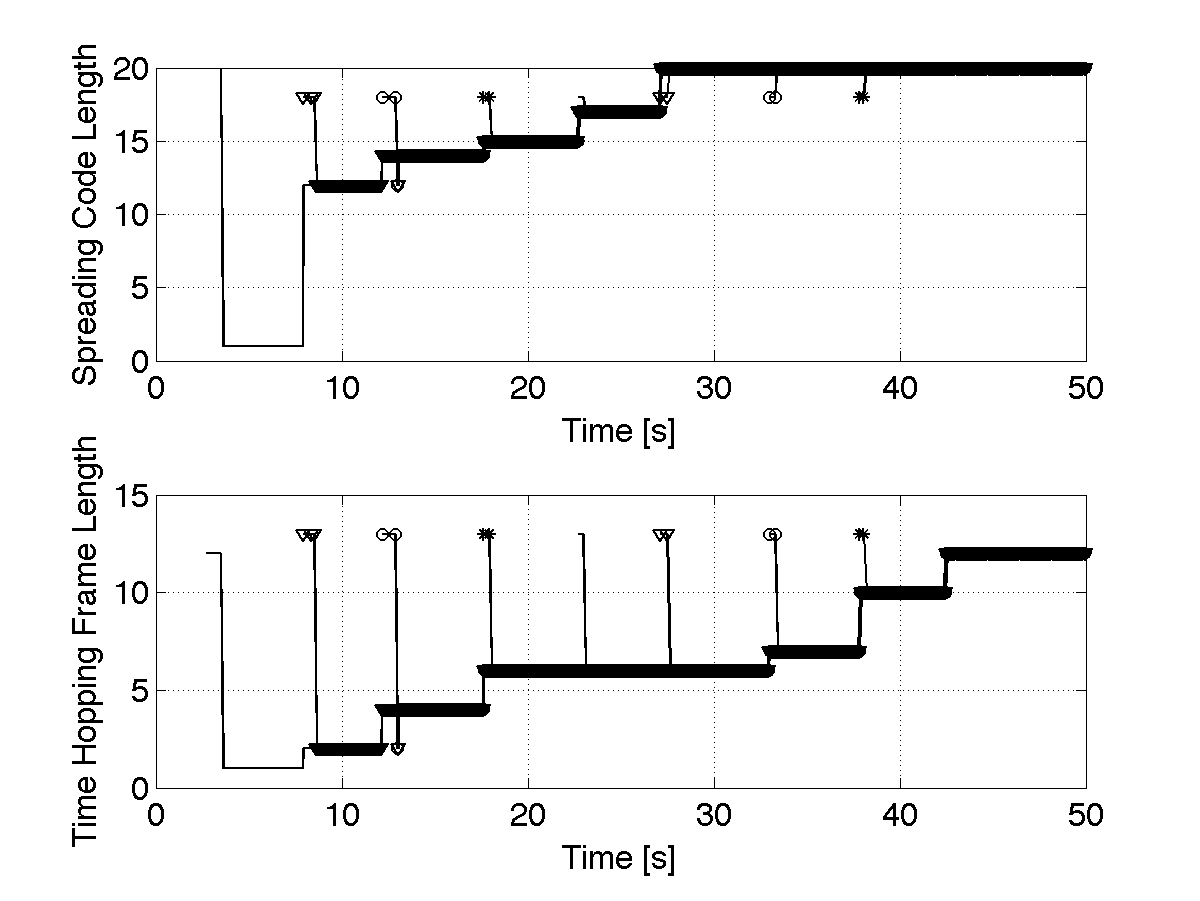}
\caption{ {\footnotesize {\bf  Time evolution of TH frame and spreading code lengths for implicitly cooperative problem.}}}
\label{framecode_conn9}
\end{figure}

We then focus on the analysis of the dynamic adaptation of the frame and code length performed distributively by each node. Focusing on the coherent system, in Fig. \ref{framecode_conn9} we show the evolution in time of frame and code length when 9 different connections are asynchronously activated with a deterministic $5\:\mathrm{s}$ delay between each other. As expected, each connection starts with the maximum supported frame and code length and then adaptively reaches the optimal value based on the interference level measured at the receiver. Since the number of interferers is the same for each receiver, the locally optimal solution is also globally optimal. 

%VERSIONE SINGOLA

\begin{figure}
\centering
\includegraphics[width=3.5in]{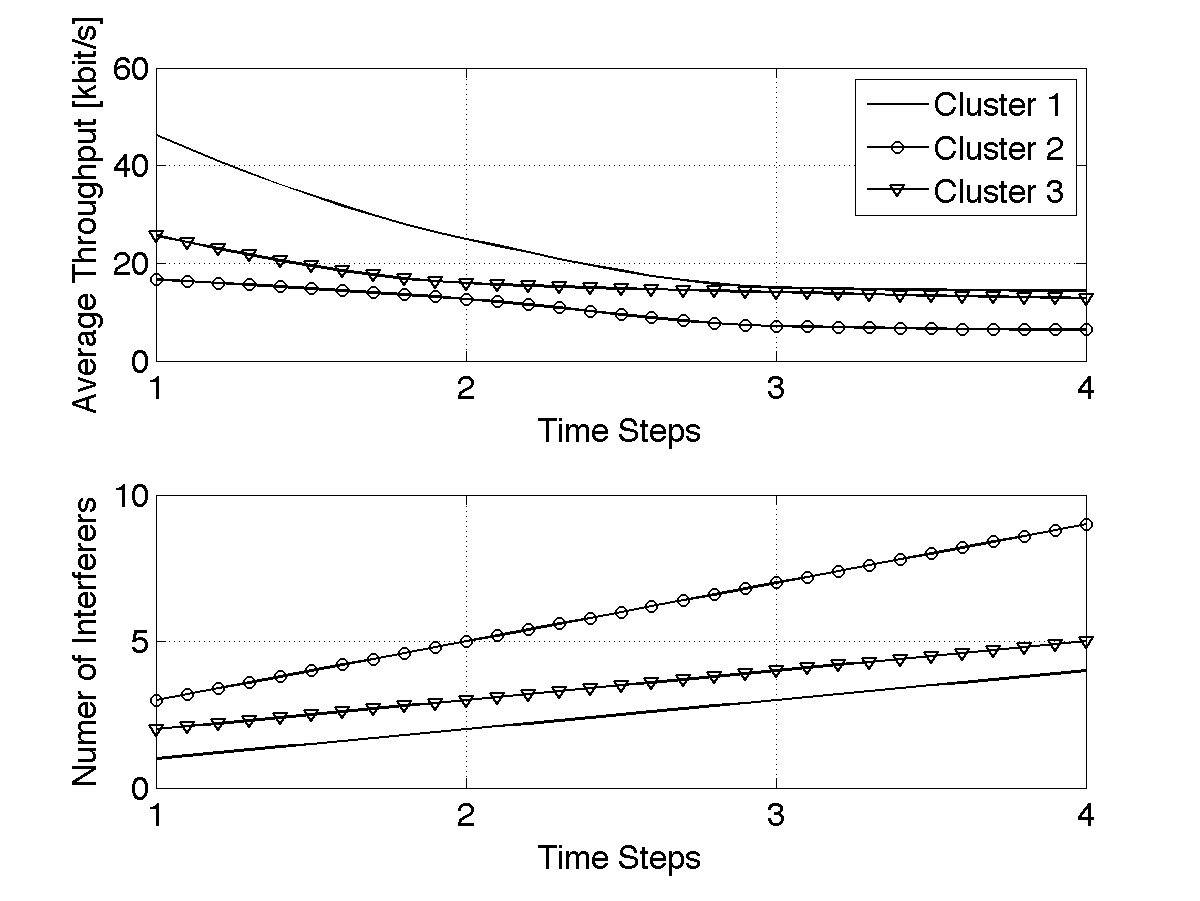} 
\caption{ {\footnotesize {\bf Average throughput [kbit/s] with explicitly cooperative problem for different number of interferers.}}}
\label{thrgpt_int_new_3clus}
\end{figure}

\begin{figure}
\centering
\includegraphics[width=3.5in]{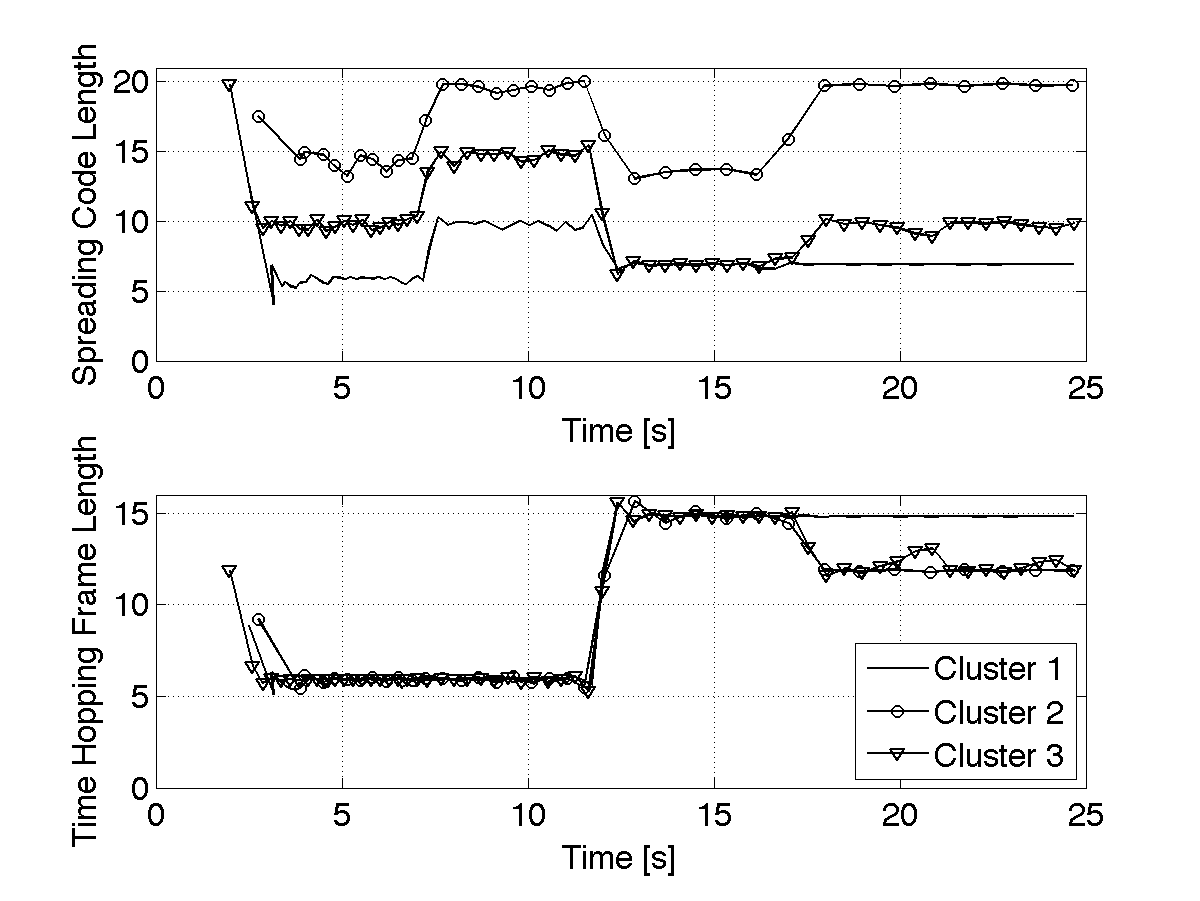}
\caption{ {\footnotesize {\bf  Time evolution of TH frame and spreading code lengths for explicitly cooperative problem.}}}
\label{frame4conn9_clu}
\end{figure}

{\bf Explicitly cooperative solution}
In Fig. \ref{thrgpt_int_new_3clus} we plot the network throughput obtained when frame and code length are adaptively regulated based on the explicitly cooperative problem in \eqref{distr_prob}. The solution is presented for the PPM-BPSK-spread transmission strategies discussed in Section \ref{uswb}. Throughput is evaluated by varying the number of active connections in four consecutive time steps. In particular, we assume that there is always one active connection in cluster \#2. Cluster \#1 and \#3, activate a new connection at each time step. In Fig. \ref{thrgpt_int_new_3clus}, we also report the average number of interferers measured by the receivers in each cluster. As expected, the cluster located in the middle, hence located within the transmission range of both the other two clusters, measures a higher level of interference, and therefore achieves a lower throughput. However, since the BER constraint is satisfied by all nodes in the three clusters, a low packet drop rate is achieved. 
Finally, in Fig. \ref{frame4conn9_clu} the dynamic behavior of the frame and code length adaptation is shown. We consider two new connections activated asynchronously every 5 s in cluster \#1 and \#3.  In each cluster, the frame and code length are adapted according to the increasing level of interference measured in the channel. In particular, we observe the effect of the constraint in (\ref{sinr_const_game4_2}), which forces the frame length to be greater or equal than the frame length of the connection that is experiencing the highest level of interferences, i.e., the connection in cluster \#2.

\subsection{Rate-optimal Vs. Energy-optimal Results}
\begin{figure}
\centering
\includegraphics[width=3.5in]{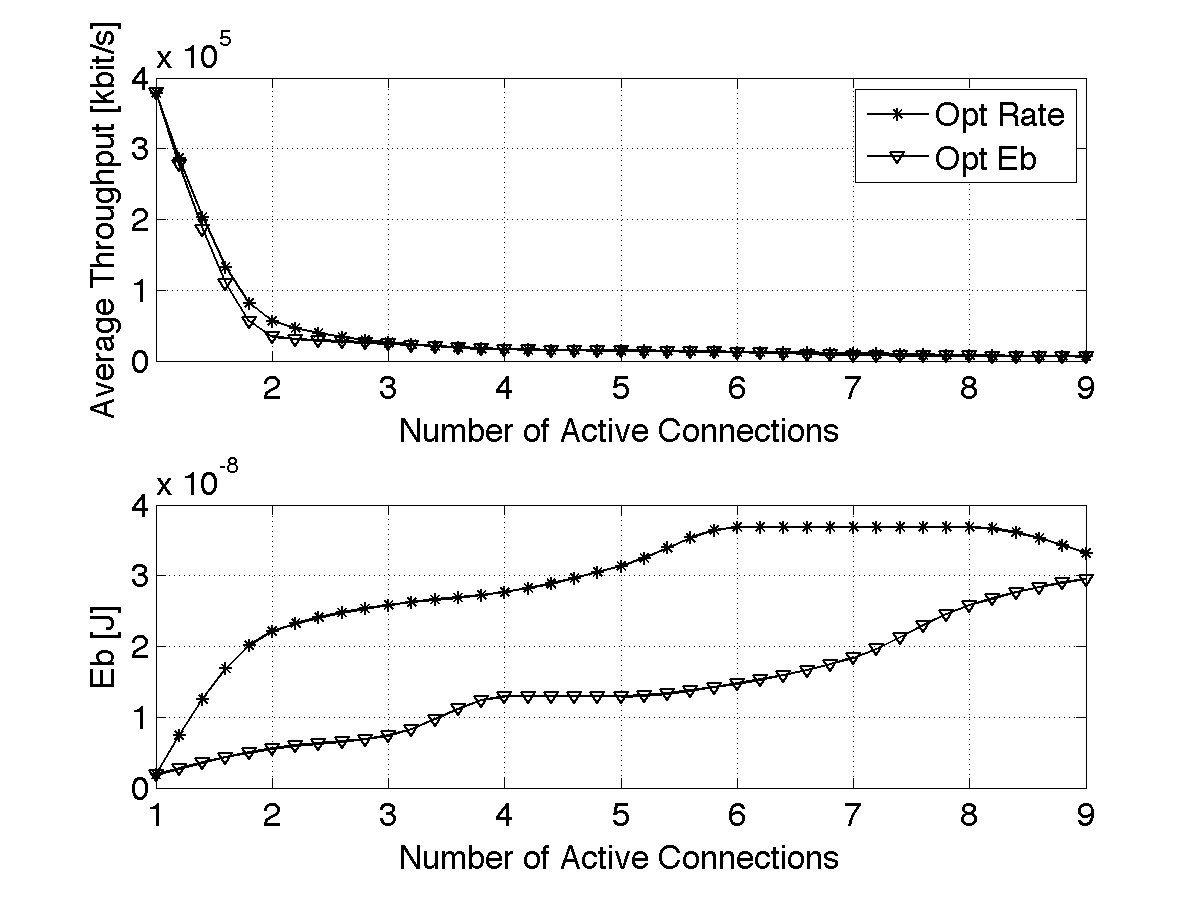}
\caption{ {\footnotesize {\bf  Throughput and $E_b$ vs. number of active connections for rate optimized and $E_b$-optimized solutions.}}}
\label{thrgpt_tot_pow9}
\end{figure}

In Figs. \ref{thrgpt_tot_pow9} we plot the throughput  obtained by adapting the transmission rate according to the energy-minimizing strategy introduced in Section \ref{sec:mac:energy}.  The energy-minimizing strategy is also compared to the rate-maximizing strategy introduced in Section \ref{sec:mac:rate} in terms of data rate and energy consumption. 
The throughput achievable in case of the rate-maximizing solution is comparable to what is obtained with the energy-optimized solution. When the number of active connections, $N$, is lower than 2, the throughput of the $E_b$-optimal strategy is close to the throughput of the rate-maximizing scheme. For $N$ higher than 2, the two considered strategies show a similar throughput performance. Since the maximum BER constraints are always satisfied, the rate-maximizing and the energy-minimizing strategies lead to packet drop rates close to zero. 
Finally, the energy consumption obtained with the $E_b$-optimal solutions, is always better that what can be obtained using the rate-maximizing scheme.

\section{Conclusions} \label{conclusions}
We proposed a paradigm shift in networking through body tissues to address
 the limitations of RF propagation in the human body.  We presented the first attempt at enabling networked intra-body communications among miniaturized sensors and actuators
 using ultrasonic waves.
 
 The contribution of our work is manyfold.  We assessed the feasibility of using ultrasonic communications within the human body;
we derived an accurate channel model for ultrasonic communications in the human body and 
 built on it to propose a new ultrasonic transmission and multiple access technique, denoted as UsWB, based on  transmission of short duration pulses following a time hopping pattern. Simulation results demonstrate the high throughput, low packet drop rate, and low energy consumption of UsWB in intra-body communications.   %%%%%%%%%%%%%%%%%

\bibliography{secon2013}{}
\bibliographystyle{unsrt}

\end{document}